\documentclass[aps,pra,twocolumn,groupedaddress,showpacs]{revtex4}
\usepackage{color}
\usepackage[colorlinks=true,linkcolor=blue,urlcolor=blue]{hyperref}	
\usepackage[singlelinecheck=false, font={rm, sl, small}, textfont={up, md}, justification=RaggedRight]{caption}
\usepackage[german,english]{babel}
\usepackage{fancyhdr}
\usepackage{amssymb,latexsym}
\usepackage{amsmath}
\usepackage[latin1]{inputenc}
\usepackage{epsfig}
\usepackage{eucal}
\usepackage{bbm}
\usepackage{amsfonts}
\usepackage{graphicx}
\begin{document}

%
%
%
%
%
\def\bra#1{\mathinner{\langle{#1}|}}
\def\ket#1{\mathinner{|{#1}\rangle}}
\def\braket#1{\mathinner{\langle{#1}\rangle}}
\def\Bra#1{\left<#1\right|}
\def\Ket#1{\left|#1\right>}
{\catcode`\|=\active 
  \gdef\Braket#1{\left<\mathcode`\|"8000\let|\BraVert {#1}\right>}}
\def\BraVert{\egroup\,\mid@vertical\,\bgroup}
%

{\catcode`\|=\active
  \gdef\set#1{\mathinner{\lbrace\,{\mathcode`\|"8000\let|\midvert #1}\,\rbrace}}
  \gdef\Set#1{\left\{\:{\mathcode`\|"8000\let|\SetVert #1}\:\right\}}}
\def\midvert{\egroup\mid\bgroup}
\def\SetVert{\egroup\;\mid@vertical\;\bgroup}

%
\begingroup
 \edef\@tempa{\meaning\middle}
 \edef\@tempb{\string\middle}
\expandafter \endgroup


\title{Controlled-NOT for multiparticle qubits and topological quantum computation based on parity measurements}


\author{Oded Zilberberg}
\author{Bernd Braunecker}
\author{Daniel Loss}
\affiliation{Department of Physics, University of Basel, Klingelbergstrasse 82, 4056 Basel, Switzerland}

\date{\today}

\begin{abstract}
We discuss a measurement-based implementation of a controlled-NOT (CNOT) quantum gate.
Such a gate has recently been discussed for free electron qubits.
Here we extend this scheme for qubits encoded in product states of two (or more) spins-1/2 or 
in equivalent systems.
The key to such an extension is to find a feasible qubit-parity meter.
We present a general scheme for reducing this qubit-parity meter to a local spin-parity measurement 
performed on two spins, one from each qubit. Two possible realizations 
of a multiparticle CNOT gate are further discussed:
electron spins in double quantum dots in the singlet-triplet encoding, and 
$\nu=5/2$ Ising non-Abelian anyons using topological quantum computation braiding operations 
and nontopological charge measurements.
\end{abstract}

\pacs{03.67.Lx, 73.21.La, 05.30.Pr, 85.35.Be}


\maketitle

\section{Introduction}
\label{intro}
Single-quantum-bit (qubit) operations and a two-qubit gate that
generates entanglement are sufficient for universal quantum computation \cite{barenco}.
One such two-qubit gate is the controlled-NOT (CNOT) which flips 
the state of a target qubit if the control qubit is in the logical $\ket{1}$ state.
A physical implementation of the CNOT gate typically requires a control of the interaction 
between the qubits, e.g. for spin qubits see \cite{divincenzo,burkard3,klauser}. However, introducing interactions
between qubits inevitably introduces additional decoherence sources and is not possible in
some quantum computation proposals such as, for instance, in linear-optics 
quantum computation due to the fact that photons interact in a negligible way.
However, Knill, Laflamme, and Milburn (KLM) have shown that measurements 
rather than interactions can provide the
means to implement a CNOT gate on photons using nonunitary operations \cite{klm}.
Shortly thereafter, additional measurement-based approaches for quantum computation 
were proposed \cite{nielsen,leung}.

The KLM model indeed served as a stepping stone for coherent quantum information processing, 
but was restricted to the underlying physical system, relying on the bosonic properties of photons. 
Attempts to design a similar implementation
for fermionic systems encountered some difficulties in the form of 
a no-go theorem \cite{terhal,knill}, which showed that for fermions, 
single-electron Hamiltonians and single-spin measurements are simulated efficiently by classical means.
This no-go theorem, however, 
was sidestepped recently in a work by Beenakker et al. \cite{beenakker}.
By taking advantage of the additional charge 
degree of freedom of an electron, a two-spin parity measurement was proposed. Using this parity meter,
a measurement-based CNOT gate for
free ``flying'' electrons was designed. Following this result, implementations of a parity
gate for spin \cite{loss} and charge qubits \cite{bjorn} 
have been proposed.

The setup in Ref. \cite{beenakker} was proposed for qubits encoded in the spin states
of free electrons, i.e. the electron spin up (down) is interpreted as a logical 1 (0) state.
Many qubit-encoding schemes, however, encode a qubit in two states from a Hilbert space
larger than the two-dimensional spin-1/2 Hilbert space, specifically, from a product
Hilbert space of two (or more) two-level systems. An example is the singlet-triplet ($S-T_0$) encoding \cite{levy}.
For simplicity, we refer to the composing particles of this kind of qubit as spins-1/2, yet
we emphasize that they can have various physical origins. 
Such encoding schemes result from system-dependent constraints, for instance, seeking a
less noisy physical system as in the case of electron spins in double quantum dots \cite{levy},
or due to topological constraints in the case of $\nu=5/2$ Ising-type anyons, 
where two quasiparticles form a two-level system equivalent to a spin-1/2 \cite{bravyi}.

In this paper we discuss a qubit-parity measurement-based implementation of a CNOT gate for such
multiparticle qubits. The implementation is a direct extension
of the schemes proposed in \cite{klm,beenakker}.
The key to such an extension is to find a feasible qubit-parity measurement.
We present a general scheme to reduce this measurement to a local spin-parity measurement 
of a representative spin from each qubit.
For concreteness, we specifically discuss qubits based on the $S-T_0$ basis and
present for this case a proof of the linearity of the measurement-based CNOT gate operation.
The linearity proof is required due to the nonlinear nature of the 
measurement-based implementation of the gate and can be used similarly
for the case of Ref. \cite{beenakker}.
We also propose a possible realization of such a $S-T_0$ CNOT on double quantum dots 
using a recently proposed spin-parity meter \cite{loss}.

For $\nu=5/2$ Ising-type anyons, a meter equivalent to a spin-parity meter involves
measuring the charge of four quasiparticles. Such measurements have been recently proposed \cite{fradkin,bks,stern,feldman,feldman2}
and first steps toward their implementation have been presented \cite{heiblum,zumbuhl}. 
In Ref. \cite{bravyi} this type of parity measurement was invoked alongside 
topological braiding operations to implement a two-qubit entangling gate.
We use this parity meter to construct the measurement-based CNOT gate for this system.
A comparison to the scheme of Ref. \cite{bravyi} shows the following differences: The present scheme 
requires only local braiding between the anyons composing a qubit but also additional
anyons for an ancilla and an additional parity measurement.
The scheme in Ref. \cite{bravyi} is thus more efficient in anyon resources and uses one
parity measurement less, but it requires long-ranged anyon braiding operations between qubits,
which will be experimentally challenging.

The paper is structured as follows: In Sec. \ref{scheme} we present the scheme for
the qubit-parity meter using a representative spin-parity measurement. The scheme
is presented fully for the $S-T_0$ qubit encoding. We then extend
the result of Ref. \cite{beenakker} and present the measurement-based CNOT setup using
the $S-T_0$ qubit-parity meter. In Sec. \ref{physical} we discuss possible implementations
of the CNOT scheme, focusing on two physically entirely different systems: double quantum dots and $\nu=5/2$ Ising-type anyons. In the Appendix
we prove the linearity of the measurement-based CNOT gate.
\section{scheme for qubit-parity measurement and CNOT}
\label{scheme}
In order to extend the measurement-based CNOT gate proposed in Ref. \cite{beenakker}
to a multiparticle qubit encoding, one must find a way to measure the qubit parity of two such qubits.
We propose a general scheme in which the qubits are rotated to ``witness'' states 
such that a representative spin parity measurement demonstrates their qubit parity. 
We illustrate this scheme on a specific two-spin singlet-triplet qubit encoding where
two selected Bell states serve as the qubit's logical state, i.e. 
$\ket{0}=\ket{T_0}=(\ket{\uparrow\downarrow}+\ket{\downarrow\uparrow})/\sqrt{2},\ket{1}=\ket{S}=(\ket{\uparrow\downarrow}-\ket{\downarrow\uparrow})/\sqrt{2}$. 
This is an encoding scheme
used for electron spins in double-quantum-dot setups \cite{levy,wu1,petta,coish,taylor,burkard2,dimitrije}.
We show that a spin-parity measurement is sufficient for a $S-T_0$ parity meter and 
detail the CNOT implementation.

In order to demonstrate the equivalence between spin parity and $S-T_0$ qubit parity, 
we rotate the qubit states to witness states over which a spin parity measurement 
will make the distinction of qubit parity. An important building block in this scheme
is the single-qubit Hadamard gate $\hat{H}$. Applied to the computational basis states $\{\ket{0},\ket{1}\}$, it has the
matrix representation
$\frac{1}{\sqrt{2}}(\begin{smallmatrix}1 && 1 \\
1 && -1\end{smallmatrix})$,
and it yields for the $S-T_0$ encoding
\begin{align}
\hat{H}\ket{T_0}=\ket{\uparrow\downarrow},\nonumber\\
\hat{H}\ket{S}=\ket{\downarrow\uparrow}.
\label{hadamard}
\end{align}

\begin{figure}
\includegraphics[width=\columnwidth]{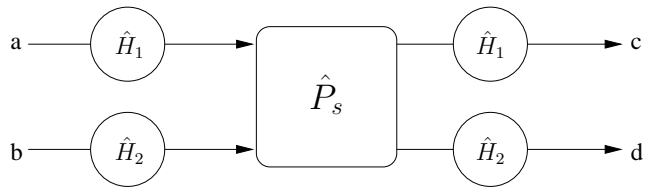}
\caption{A gate that uses a spin-parity measurement to measure the parity of $S-T_0$ qubits. 
A pair of $S-T_0$ qubits enters the gate in arms $a$ and $b$. Each of the qubits is rotated by a 
Hadamard gate $\hat{H}$. The spin parity of the left spins (see Eq. \eqref{hadamard}) from each qubit
is then measured in the $\hat{P}_s$ box. The qubits are rotated back by 
Hadamard gates and the parity of the spins is 
equivalent to the parity of the qubits.}
\label{parity}
\end{figure}

Therefore, the left spin in the right-hand side of Eq. \eqref{hadamard} 
can serve as a witness for the original 
two-spin state. For example, if the left spin is in the $\ket{\uparrow}$ state, the original 
prerotated state was a $\ket{T_0}$.
Hence, the spin parity of the left spins of two rotated $S-T_0$ qubits indicates
the qubit parity. If $\hat{P}_s$ is a spin-parity gate 
(as used in Refs. \cite{beenakker,radu}) we obtain a $S-T_0$ qubit-parity gate from the operation
\begin{equation}
\hat{P}=\hat{H}_1\hat{H}_2\hat{P}_s\hat{H}_1\hat{H}_2,
\label{parity_op}
\end{equation}
where $\hat{H}_1,\hat{H}_2$ are the Hadamard gates operating on qubits $1$ and $2$,
and $\hat{P}_s$ measures the spin parity between the two left spins of qubits $1$ and $2$.
A sketch of this gate is shown in Fig. \ref{parity}.

As an example for the operation of $\hat{P}$, 
let $\ket{\psi}=\ket{T_0}_1\otimes(\alpha\ket{T_0}_2+\beta\ket{S}_2)$ 
be a two-qubit state. Once rotated by Hadamard gates the state becomes
$\ket{\tilde{\psi}}=\ket{\uparrow\downarrow}_1\otimes(\alpha\ket{\uparrow\downarrow}_2+\beta\ket{\downarrow\uparrow}_2)$.
Measuring the spin parity of the left spins in each qubit results in
$\ket{\tilde{\psi}_{\{1\}}}=\ket{\uparrow\downarrow}_1\otimes\ket{\uparrow\downarrow}_2$
if even spin parity is measured ($p_s=1$) and 
$\ket{\tilde{\psi}_{\{0\}}}=\ket{\uparrow\downarrow}_1\otimes\ket{\downarrow\uparrow}_2$
if odd spin parity is measured ($p_s=0$). Rotating the qubits by Hadamard gates 
again results in the projected qubit states with a qubit parity equivalent to the measured spin parity.

The fact that Hadamard gates rotate to witness states and back in this $S-T_0$ encoding
results from the fact that the computational states are a superposition of the two product spin states
$\{\ket{\uparrow\downarrow},\ket{\downarrow\uparrow}\}$ with equal amplitudes.
Thus, for an x-aligned single-spin qubit encoding 
$\ket{\pm}=(\ket{\uparrow}\pm\ket{\downarrow})/1/\sqrt{2}$ the same
qubit parity routine as shown in Fig. \ref{parity} is valid.
The difference between the above setup and a parity setup for other types of encoding 
lies in the single-qubit rotation
that rotates the qubits to the witness states and back, i.e. the 
Hadamard gates in the entrance and exit of the parity gate are replaced by different rotations 
for other encoding schemes.
Additionally, one must note that some of such rotations to witness states might require
the qubit to leave the computational subspace. For example, if one uses a unitary two-spin
rotation that maps 
$\ket{S}\rightarrow\ket{\downarrow\downarrow},\ket{T_0}\rightarrow\ket{\uparrow\uparrow}$,
the target states are outside the qubit encoding subspace. However, they can still be 
used as witness states
for the qubit-parity measurement. In this case it is, of course, required that those additional
states are energetically accessible from the computational subspace.

\begin{figure}
\includegraphics[width=\columnwidth]{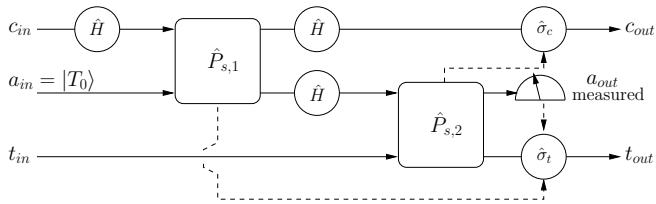}
\caption{Measurement-based CNOT gate for $S-T_0$ qubits. The boxes represent spin-parity 
measurements of the left spins (see Eq. \eqref{hadamard}) of each qubit. 
Three Hadamard gates rotate the qubits entering and leaving the first box. The input of 
the CNOT gate consists of control and target qubits plus an ancilla which is prepared in the $\ket{T_0}$ state. 
The ancilla is measured at the output in a $\ket{S}$ or $\ket{T_0}$ state. The outcome of this 
measurement plus the two measured spin parities determine which operators $\hat{\sigma}_c,\hat{\sigma}_t$ one 
has to apply on the control and target qubits, respectively, in order to complete the CNOT operation: 
We apply on the control qubit
$\hat{\sigma}_c=\hat{\sigma}_z$ if $p_2=0$ and $\hat{\sigma}_c=\mathbf{1}$ if $p_2=1$. For the target qubit, 
$\hat{\sigma}_t=\hat{\sigma}_x$ if $p_1=1$ and the ancilla is measured in the $\ket{S}$ state, or if $p_1=0$ 
and the ancilla is measured in the $\ket{T_0}$ state. Otherwise, $\hat{\sigma}_t=\mathbf{1}$. See \cite{beenakker}.}
\label{gate}
\end{figure}

We can now extend the result of Ref. \cite{beenakker}, where spin parity is used 
to implement a CNOT operation on a single-spin qubit, by using the $S-T_0$ parity 
in the same way for a $S-T_0$ CNOT implementation.
The resulting gate is shown in Fig. \ref{gate}.
The gate can be seen as a Hadamard-rotated version of the gate from Ref. \cite{beenakker}
that operates on two-spin qubits instead of free flying electron qubits. 
In addition, the gate has the following advantages over the gate from Ref. \cite{beenakker}:
(1) The ancilla is prepared in a pure computational state instead of a 
superposition of computational states, and
(2) fewer Hadamard operations are required.

\begin{figure*}
\includegraphics[width=\textwidth]{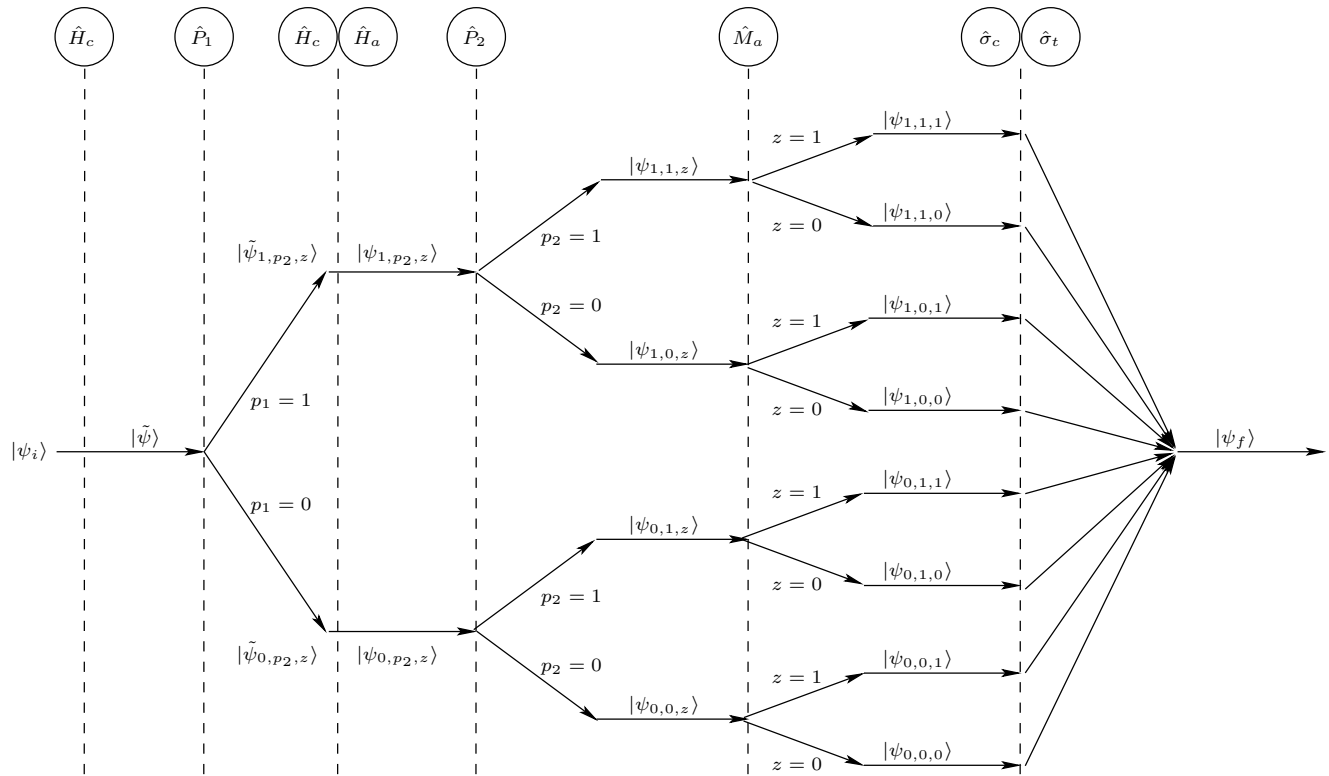}
\caption{Calculation tree of the measurement-based CNOT gate in Fig. \ref{gate}. The computation splits in 
accordance with the parity and ancilla measurements. Due to the Hadamard gate rotations, 
the measurements do not destroy the initial state and each path of the computation has 
the same probability of occurring. We obtain eight possible result states which we denote 
as $\ket{\psi_{p_1,p_2,z}}$. In the Appendix we follow the execution of the gate and show that the results in each 
of the calculation arms indeed merge into a single result $\ket{\psi_f}$ which is equal to the result 
of the CNOT operation.}
\label{comp_route}
\end{figure*}

The parity and ancilla measurements (see Fig. \ref{gate}) are projective nonlinear operations. Each measurement 
projects the state onto one of two possible outcome states. In Fig. \ref{comp_route} we present the ``calculation tree''
of the CNOT gate where the three consecutive measurements lead to
eight possible outcome states. With the last tuning step of the 
gate, however, we obtain a single deterministic result, i.e. all branches have the outcome (up to a global phase),
\begin{align}
\ket{\psi}_c\otimes\ket{\psi}_t\rightarrow&\alpha\gamma\ket{T_0}_c\ket{T_0}_t+\alpha\delta\ket{T_0}_c\ket{S}_t+\nonumber\\
&\beta\gamma\ket{S}_c\ket{S}_t+\beta\delta\ket{S}_c\ket{T_0}_t,
\label{cnot_res}
\end{align}
where $\ket{\psi}_c=\alpha\ket{T_0}_c+\beta\ket{S}_c$ and 
$\ket{\psi}_t=\gamma\ket{T_0}_t+\delta\ket{S}_t$ are the control and target input states, respectively. Hence Eq. \eqref{cnot_res} describes
the operation of a CNOT gate on $S-T_0$ qubits.

In the Appendix we follow the calculation tree in Fig. \ref{comp_route} 
when the gate in Fig. \ref{gate} is applied to an 
arbitrary two-qubit state. The result yields Eq. \eqref{cnot_res} and proves that the gate is indeed a CNOT gate
and that its operation is linear.
\section{Possible implementations}
\label{physical}
We present here two possible implementations of the CNOT gate for two types 
of systems that have been proposed for quantum computation.
In the first part we discuss how it may be realized on double-quantum-dot qubits.
In the second part we consider an implementation for non-Abelian Ising-type anyons
that have been proposed to exist as elementary excitations in a fractional quantum Hall
system with filling factor $\nu=5/2$.
\subsection{Double quantum dots}
\label{dqd_sec}
Since the introduction of electron spins in quantum dots (QDs)
as a platform for quantum information processing \cite{divincenzo}, there has been much research in this direction.
Several proposals specifically focus on a $S-T_0$ qubit encoding where
two electrons in neighboring QDs form the $\ket{S}$ and $\ket{T_0}$ states.
Possible implementations of single-qubit operations as well as a CNOT gate
based on control of the design and the interactions in the system have been discussed
in the last few years \cite{levy,wu1,petta,coish,taylor,burkard2,dimitrije}.

\begin{figure}
\includegraphics[width=\columnwidth]{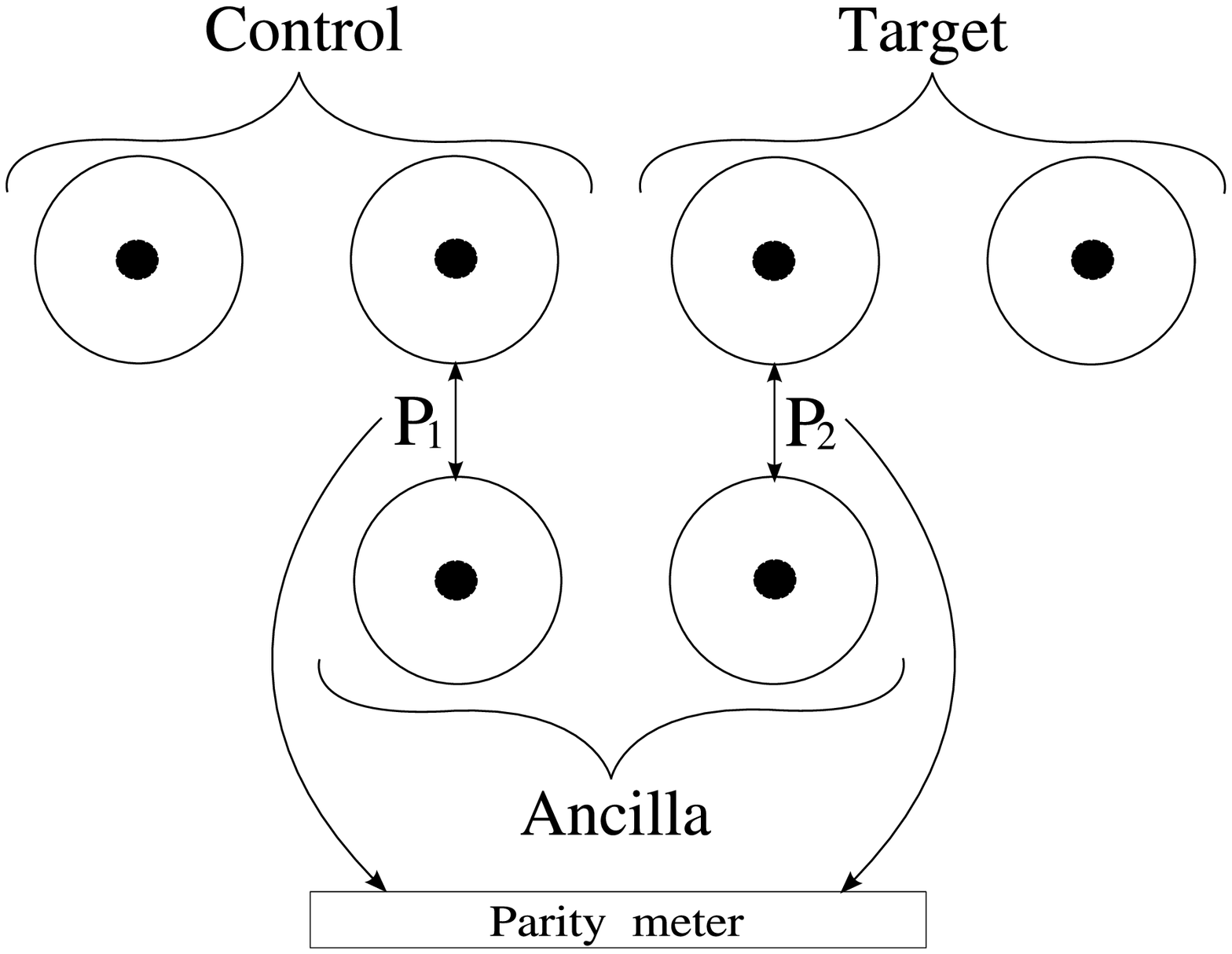}
\caption{Double-quantum-dot implementation setup for the measurement-based CNOT gate. 
A dot with an electron in it is represented by an empty circle containing a filled circle. The ancilla dots are
situated next to a spin-parity meter (proposed in Ref. \cite{loss}). 
In order to measure the parity of the ancilla and control qubits, $P_1$, the spin parity of the right
electron spin of the control and the left electron spin of the ancilla is measured.
The parity of the ancilla and target qubits, $P_2$, is measured by the spin parity of the right
electron spin of the ancilla and the left electron spin of the target.}
\label{dqd}
\end{figure}

In Fig. \ref{dqd} we show that the measurement-based CNOT gate can be realized in such systems as well.
The spin parity can be measured using a recently proposed spin-parity meter \cite{loss}.
This meter, however, is local and cannot measure
spins in distant QDs, i.e. if we label the electrons by 1,2 (first qubit) and 3,4 (second qubit), 
the required witness parity of spins 1 and 3 cannot be measured.
From Eq. \eqref{hadamard} we see, however, that measuring even parity between spins 2 and 3 is
the same as measuring odd parity between spins 1 and 3, and vice versa. Upon this reinterpretation,
the CNOT gate remains unchanged.
With the suggested geometric arrangement of QDs shown in Fig. \ref{dqd}, it may further
be possible that a single spin-parity meter, coupling alternately to the left or right
QD of the ancilla, is sufficient for the operation.
\subsection{Ising-type anyons}
\label{anyon_sec}
Topological quantum computation (TQC) \cite{kitaev1,kitaev2,kitaev3} proposes a scheme
in which coherent quantum computation is done by topological operations performed on 
non-Abelian anyons. 
A physical system that may serve as a platform for TQC is the two-dimensional electron gas
in the fractional quantum Hall regime. At filling fraction $\nu=5/2$, localized
elementary excitations (quasiparticles) are proposed to have non-Abelian anyon statistics and are dubbed 
Ising anyons \cite{moore,nayak,freedman}. 

Two such quasiparticles form a two-level system equivalent to a spin-1/2.
However, in Ref. \cite{bravyi} it is shown that due to topological superselection rules 
the qubit is encoded in two product states $\ket{0}=\ket{0,0},\ket{1}=\ket{1,1}$ 
from the Hilbert space formed by four quasiparticles. Thus, this system forms a Hilbert space 
equivalent to that of a product Hilbert space of two spins-1/2 \footnote{An alternative qubit encoding scheme to the four-quasiparticle qubit is to encode the qubit in three quasiparticles, for which a CNOT gate has been proposed in \cite{georgiev}.}.
In addition, it is shown in Ref. \cite{bravyi} that in order to implement universal quantum computation on this system, 
nontopological parity-like measurements are required. Such measurements may be carried out by an interferometric 
device recently proposed in \cite{fradkin,bks,stern,feldman,feldman2} and first steps have been taken
toward its implementation \cite{heiblum,zumbuhl}. We refer to these references for more details.

\begin{figure}
\includegraphics[width=6cm]{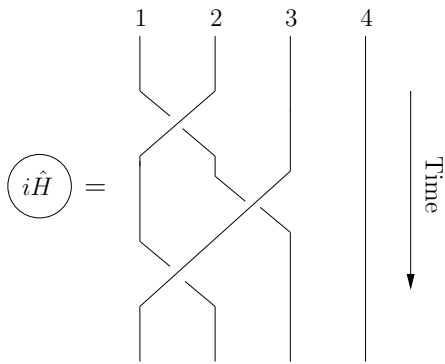}
\caption{Hadamard gate using braiding of $\nu=5/2$ Ising anyons \cite{georgiev,zhang}. The gate adds a global $\pi/2$ phase which can be ignored.}
\label{anyon-h}
\end{figure}

The measurement-based CNOT scheme can be implemented over this system as well. 
The parity meter here acts directly on the computational states so that no rotation
prior to the parity measurement is required. If we label the anyons forming the
first qubit 1,2,3,4 and those of the second qubit 5,6,7,8, the parity of two qubits can 
be measured by an interferometer which measures the charge of the four adjacent anyons 3,4,5,6.
This measurement is equivalent to the spin-parity measurement of two neighboring spins, one from each qubit, as discussed in Sec. \ref{dqd_sec}.
The required Hadamard rotations by the measurement-based CNOT scheme \cite{beenakker}
can be implemented using topological braiding of the Ising anyons \cite{georgiev,zhang}.
If we consider the qubit formed by the anyons 1,2,3,4, the braiding of anyons 1,2 results in a
$e^{i\frac{\pi}{4}\hat{\sigma}_z}$ qubit rotation and braiding of 2,3 results in
$e^{i\frac{\pi}{4}\hat{\sigma}_x}$ \cite{bravyi}. Since 
$i\hat{H}=e^{i\frac{\pi}{4}\hat{\sigma}_z}e^{i\frac{\pi}{4}\hat{\sigma}_x}e^{i\frac{\pi}{4}\hat{\sigma}_z}$, 
the operation shown in Fig. \ref{anyon-h}
performs a Hadamard rotation on such a qubit (up to a global phase) \cite{georgiev,zhang}.

\begin{figure} 
\includegraphics[width=\columnwidth]{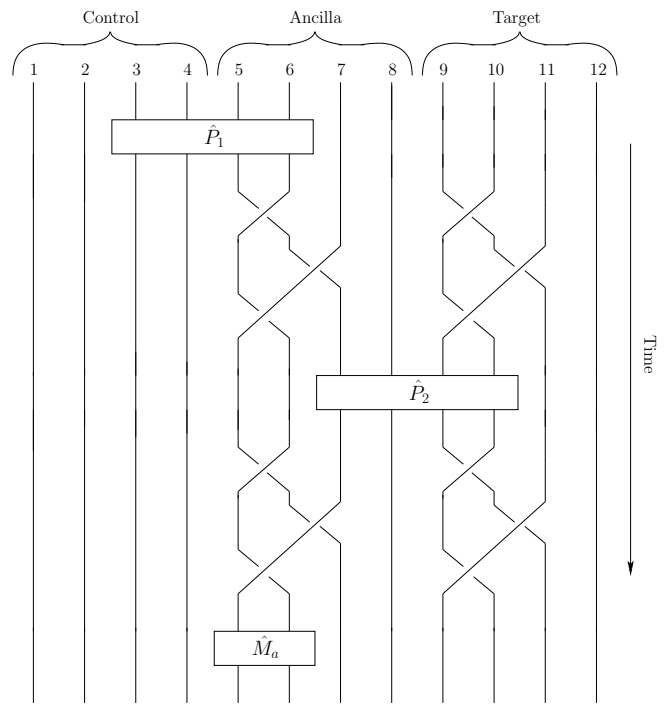}
\caption{Measurement-based CNOT gate implemented on $\nu=5/2$ Ising anyon qubits.
The control, ancilla, and target qubits are shown from left to right,
e.g. the control qubit is represented by anyons 1,2,3,4. The representative
``spin''-parity measurements are shown by the $\hat{P}$ boxes and the ancilla
measurement by the box at the bottom. The braiding between the measurements 
represents Hadamard rotations on the qubits.}
\label{anyon-cnot}
\end{figure}

Figure \ref{anyon-cnot} presents the measurement-based
CNOT operation on this system. The parity measurement of anyons 3,4,5,6 gives the parity
of the control and ancilla qubits. The parity measurement of anyons 7,8,9,10 gives the parity
of the ancilla and target qubits. 

Upon comparing the CNOT operation in Fig. \ref{anyon-cnot} to the two-qubit gate proposed in Ref. \cite{bravyi},
we see that the present measurement-based scheme requires an ancilla and an additional parity measurement
that are not needed in \cite{bravyi}. This is due to
the fact that the present scheme does not take advantage of the underlying anyon system statistics.
Using the anyon statistics as in Ref. \cite{bravyi}, however,
requires braiding of distant anyons between qubits, while in the present scheme all braiding operations 
are strictly local, i.e. we require only braiding operations of nearest-neighbor quasiparticles.
Furthermore, we braid the anyons only within the qubit they define.
\section{Conclusion}
We have presented a general scheme to measure the qubit parity of two multiparticle qubits via a 
representative spin-parity measurement in some rotated state.
Using this qubit-parity meter we have extended the measurement-based CNOT setup proposed  
in Ref. \cite{beenakker} to additional encoding schemes.
As an example, we discussed the $S-T_0$ qubit encoding case in detail. In this encoding, 
as shown in Fig. \ref{gate}, 
the rotations used by the qubit-parity meter led to a slightly simpler 
rotated setup of the CNOT gate as compared to \cite{beenakker}.
We also used the $S-T_0$ setup to provide a proof of the linearity of the gate 
(see Fig. \ref{comp_route}), which is required as the CNOT is implemented by nonunitary operations.
As an illustration, we presented two possible implementations of the CNOT gate.
We have proposed a possible setup for the $S-T_0$ encoding (see Fig. \ref{dqd}).
For $\nu=5/2$ Ising-type anyons, the CNOT gate can be implemented with braiding operations
and the parity meter proposed in \cite{bravyi} (see Fig. \ref{anyon-cnot}).
In contrast to a similar gate described in \cite{bravyi}, the present CNOT
requires one more parity measurement and the additional ancilla.
But all braiding operations remain strictly local and confined within the individual qubits.
Both schemes have their strengths but it is yet unknown which is a more efficient route
for implementation.
\begin{acknowledgments}
We thank W. A. Coish, S. Bravyi, L. Chirolli, D. Stepanenko, and D. Zumb{\"u}hl for useful discussions.
Financial support by the NCCR Nanoscience and the Swiss NSF is acknowledged.
\end{acknowledgments}
\appendix
\section{Proof of linearity}
\label{linearproof}
To prove the linearity of the measurement-based CNOT gate shown in Fig. \ref{gate},
we follow the gate execution that is portrayed in Fig. \ref{comp_route} when the
control and target qubits are taken initially to be in arbitrary states:
\begin{align}
\ket{c}=\alpha\ket{0}+\beta\ket{1}=\alpha\ket{T_0}+\beta\ket{S}\\
\ket{t}=\gamma\ket{0}+\delta\ket{1}=\gamma\ket{T_0}+\delta\ket{S}
\end{align}

The initial state of the input qubits plus the ancilla is:
\begin{eqnarray}
\ket{\psi_i}=\ket{c}\otimes\ket{a}\otimes\ket{t}.
\end{eqnarray}

The calculation splits in accordance with the parity and ancilla measurements. We obtain
eight optional result states which we denote as $\ket{\psi_{p_1,p_2,z}}$ 
as seen in Fig. \ref{comp_route}. We prove that the results in each of the calculation 
arms finally merge into a single result $\ket{\psi_f}$ which is equal to the result of the CNOT operation.

At the first step, the control is passed through a $S-T_0$ Hadamard gate, resulting in
\begin{eqnarray}
\ket{\tilde{\psi}}=&\left[\frac{\alpha}{\sqrt{2}}(\ket{T_0}_c+\ket{S}_c)+\frac{\beta}{\sqrt{2}}(\ket{T_0}_c-\ket{S}_c)\right]\otimes\nonumber\\
&\ket{a}\otimes\ket{t}.
\end{eqnarray} 

The first parity measurement is performed on the control and ancilla qubits. To present
the result of the spin-parity measurement, we first write the control and ancilla in the product spin basis $\{\ket{\uparrow\downarrow},\ket{\downarrow\uparrow}\}$,
\begin{eqnarray}
\ket{\tilde{\psi}}=\left[\alpha\ket{\uparrow\downarrow}_c+\beta\ket{\downarrow\uparrow}_c\right]\otimes\frac{\ket{\uparrow\downarrow}_a+\ket{\downarrow\uparrow}_a}{\sqrt{2}}\otimes \ket{t}.
\end{eqnarray}

Measurement of the spin parity of the left spins of the control and ancilla qubits has two possible outcomes:
\begin{align}
\ket{\tilde{\psi}_{1,p_2,z}}=&\left[\alpha\ket{\uparrow\downarrow}_c\ket{\uparrow\downarrow}_a+\beta\ket{\downarrow\uparrow}_c\ket{\downarrow\uparrow}_a)\right]\otimes\ket{t},\\
\ket{\tilde{\psi}_{0,p_2,z}}=&\left[\alpha\ket{\uparrow\downarrow}_c\ket{\downarrow\uparrow}_a+\beta\ket{\downarrow\uparrow}_c\ket{\uparrow\downarrow}_a\right]\otimes\ket{t}.
\end{align}

The ancilla and control qubits are then rotated by Hadamard gates:
\begin{align}
\ket{\psi_{1,p_2,z}}=\left[\alpha\ket{T_0}_c\ket{T_0}_a+\beta\ket{S}_c\ket{S}_a\right]\otimes\ket{t},\\
\ket{\psi_{0,p_2,z}}=\left[\alpha\ket{T_0}_c\ket{S}_a+\beta\ket{S}_c\ket{T_0}_a\right]\otimes\ket{t}.
\end{align}

Now the ancilla and target qubits enter a spin-parity measurement. Once more we write their states in the product spin basis:
\begin{align}
\ket{\psi_{1,p_2,z}}=&\left[\alpha\ket{T_0}_c\frac{\ket{\uparrow\downarrow}_a+\ket{\downarrow\uparrow}_a}{\sqrt{2}}+\beta\ket{S}_c\frac{\ket{\uparrow\downarrow}_a-\ket{\downarrow\uparrow}_a}{\sqrt{2}}\right]\otimes\nonumber\\
& \left[\gamma\frac{\ket{\uparrow\downarrow}_t+\ket{\downarrow\uparrow}_t}{\sqrt{2}}+\delta\frac{\ket{\uparrow\downarrow}_t-\ket{\downarrow\uparrow}_t}{\sqrt{2}}\right],\\
\ket{\psi_{0,p_2,z}}=&\left[\alpha\ket{T_0}_c\frac{\ket{\uparrow\downarrow}_a-\ket{\downarrow\uparrow}_a}{\sqrt{2}}+\beta\ket{S}_c\frac{\ket{\uparrow\downarrow}_a+\ket{\downarrow\uparrow}_a}{\sqrt{2}}\right]\otimes\nonumber\\
& \left[\gamma\frac{\ket{\uparrow\downarrow}_t+\ket{\downarrow\uparrow}_t}{\sqrt{2}}+\delta\frac{\ket{\uparrow\downarrow}_t-\ket{\downarrow\uparrow}_t}{\sqrt{2}}\right].
\end{align}

Measuring the spin parity of the left spins of the ancilla and target qubits again splits the result set into two possible branches:
\begin{align}
\ket{\psi_{1,1,z}}=&\frac{\alpha\gamma}{\sqrt{2}}\left[\ket{T_0}_c\ket{\uparrow\downarrow}_a\ket{\uparrow\downarrow}_t+\ket{T_0}_c\ket{\downarrow\uparrow}_a\ket{\downarrow\uparrow}_t\right]+\nonumber\\
&\frac{\alpha\delta}{\sqrt{2}}\left[\ket{T_0}_c\ket{\uparrow\downarrow}_a\ket{\uparrow\downarrow}_t-\ket{T_0}_c\ket{\downarrow\uparrow}_a\ket{\downarrow\uparrow}_t\right]+\nonumber\\
&\frac{\beta\gamma}{\sqrt{2}}\left[\ket{S}_c\ket{\uparrow\downarrow}_a\ket{\uparrow\downarrow}_t-\ket{S}_c\ket{\downarrow\uparrow}_a\ket{\downarrow\uparrow}_t\right]+\nonumber\\
&\frac{\beta\delta}{\sqrt{2}}\left[\ket{S}_c\ket{\uparrow\downarrow}_a\ket{\uparrow\downarrow}_t+\ket{S}_c\ket{\downarrow\uparrow}_a\ket{\downarrow\uparrow}_t\right],
\label{secondparity1}
\end{align}
\begin{align}
\ket{\psi_{1,0,z}}=&\frac{\alpha\gamma}{\sqrt{2}}\left[\ket{T_0}_c\ket{\uparrow\downarrow}_a\ket{\downarrow\uparrow}_t+\ket{T_0}_c\ket{\downarrow\uparrow}_a\ket{\uparrow\downarrow}_t\right]-\nonumber\\
&\frac{\alpha\delta}{\sqrt{2}}\left[\ket{T_0}_c\ket{\uparrow\downarrow}_a\ket{\downarrow\uparrow}_t-\ket{T_0}_c\ket{\downarrow\uparrow}_a\ket{\uparrow\downarrow}_t\right]+\nonumber\\
&\frac{\beta\gamma}{\sqrt{2}}\left[\ket{S}_c\ket{\uparrow\downarrow}_a\ket{\downarrow\uparrow}_t-\ket{S}_c\ket{\downarrow\uparrow}_a\ket{\uparrow\downarrow}_t\right]-\nonumber\\
&\frac{\beta\delta}{\sqrt{2}}\left[\ket{S}_c\ket{\uparrow\downarrow}_a\ket{\downarrow\uparrow}_t+\ket{S}_c\ket{\downarrow\uparrow}_a\ket{\uparrow\downarrow}_t\right],
\label{secondparity2}
\end{align}
\begin{align}
\ket{\psi_{0,1,z}}=&\frac{\alpha\gamma}{\sqrt{2}}\left[\ket{T_0}_c\ket{\uparrow\downarrow}_a\ket{\uparrow\downarrow}_t-\ket{T_0}_c\ket{\downarrow\uparrow}_a\ket{\downarrow\uparrow}_t\right]+\nonumber\\
&\frac{\alpha\delta}{\sqrt{2}}\left[\ket{T_0}_c\ket{\uparrow\downarrow}_a\ket{\uparrow\downarrow}_t+\ket{T_0}_c\ket{\downarrow\uparrow}_a\ket{\downarrow\uparrow}_t\right]+\nonumber\\
&\frac{\beta\gamma}{\sqrt{2}}\left[\ket{S}_c\ket{\uparrow\downarrow}_a\ket{\uparrow\downarrow}_t+\ket{S}_c\ket{\downarrow\uparrow}_a\ket{\downarrow\uparrow}_t\right]+\nonumber\\
&\frac{\beta\delta}{\sqrt{2}}\left[\ket{S}_c\ket{\uparrow\downarrow}_a\ket{\uparrow\downarrow}_t-\ket{S}_c\ket{\downarrow\uparrow}_a\ket{\downarrow\uparrow}_t\right],
\label{secondparity3}
\end{align}
\begin{align}
\ket{\psi_{0,0,z}}=&\frac{\alpha\gamma}{\sqrt{2}}\left[\ket{T_0}_c\ket{\uparrow\downarrow}_a\ket{\downarrow\uparrow}_t-\ket{T_0}_c\ket{\downarrow\uparrow}_a\ket{\uparrow\downarrow}_t\right]-\nonumber\\
&\frac{\alpha\delta}{\sqrt{2}}\left[\ket{T_0}_c\ket{\uparrow\downarrow}_a\ket{\downarrow\uparrow}_t+\ket{T_0}_c\ket{\downarrow\uparrow}_a\ket{\uparrow\downarrow}_t\right]+\nonumber\\
&\frac{\beta\gamma}{\sqrt{2}}\left[\ket{S}_c\ket{\uparrow\downarrow}_a\ket{\downarrow\uparrow}_t+\ket{S}_c\ket{\downarrow\uparrow}_a\ket{\uparrow\downarrow}_t\right]-\nonumber\\
&\frac{\beta\delta}{\sqrt{2}}\left[\ket{S}_c\ket{\uparrow\downarrow}_a\ket{\downarrow\uparrow}_t-\ket{S}_c\ket{\downarrow\uparrow}_a\ket{\uparrow\downarrow}_t\right].
\label{secondparity4}
\end{align}

The ancilla and target qubits can be rewritten in the $S-T_0$ basis using the following relations:
\begin{align}
\ket{\uparrow\downarrow}=\frac{1}{\sqrt{2}}\left[\ket{T_0}+\ket{S}\right]\nonumber\\
\ket{\downarrow\uparrow}=\frac{1}{\sqrt{2}}\left[\ket{T_0}-\ket{S}\right]
\label{basisspan}
\end{align}

Substituting Eq. \eqref{basisspan} into Eqs. \eqref{secondparity1}-\eqref{secondparity4} and rewriting the expressions we obtain:
\begin{align}
\ket{\psi_{1,1,z}}=&\frac{\alpha\gamma}{\sqrt{2}}\left[\ket{T_0}_c\ket{T_0}_a\ket{T_0}_t+\ket{T_0}_c\ket{S}_a\ket{S}_t\right]+\nonumber\\
&\frac{\alpha\delta}{\sqrt{2}}\left[\ket{T_0}_c\ket{S}_a\ket{T_0}_t+\ket{T_0}_c\ket{T_0}_a\ket{S}_t\right]+\nonumber\\
&\frac{\beta\gamma}{\sqrt{2}}\left[\ket{S}_c\ket{S}_a\ket{T_0}_t+\ket{S}_c\ket{T_0}_a\ket{S}_t\right]+\nonumber\\
&\frac{\beta\delta}{\sqrt{2}}\left[\ket{S}_c\ket{T_0}_a\ket{T_0}_t+\ket{S}_c\ket{S}_a\ket{S}_t\right],
\end{align}
\begin{align}
\ket{\psi_{1,0,z}}=&\frac{\alpha\gamma}{\sqrt{2}}\left[\ket{T_0}_c\ket{T_0}_a\ket{T_0}_t-\ket{T_0}_c\ket{S}_a\ket{S}_t\right]+\nonumber\\
&\frac{\alpha\delta}{\sqrt{2}}\left[\ket{T_0}_c\ket{T_0}_a\ket{S}_t-\ket{T_0}_c\ket{S}_a\ket{T_0}_t\right]+\nonumber\\
&\frac{\beta\gamma}{\sqrt{2}}\left[\ket{S}_c\ket{S}_a\ket{T_0}_t-\ket{S}_c\ket{T_0}_a\ket{S}_t\right]+\nonumber\\
&\frac{\beta\delta}{\sqrt{2}}\left[\ket{S}_c\ket{S}_a\ket{S}_t-\ket{S}_c\ket{T_0}_a\ket{T_0}_t\right],
\end{align}
\begin{align}
\ket{\psi_{0,1,z}}=&\frac{\alpha\gamma}{\sqrt{2}}\left[\ket{T_0}_c\ket{S}_a\ket{T_0}_t+\ket{T_0}_c\ket{T_0}_a\ket{S}_t\right]+\nonumber\\
&\frac{\alpha\delta}{\sqrt{2}}\left[\ket{T_0}_c\ket{T_0}_a\ket{T_0}_t+\ket{T_0}_c\ket{S}_a\ket{S}_t\right]+\nonumber\\
&\frac{\beta\gamma}{\sqrt{2}}\left[\ket{S}_c\ket{T_0}_a\ket{T_0}_t+\ket{S}_c\ket{S}_a\ket{S}_t\right]+\nonumber\\
&\frac{\beta\delta}{\sqrt{2}}\left[\ket{S}_c\ket{S}_a\ket{T_0}_t+\ket{S}_c\ket{T_0}_a\ket{S}_t\right],
\end{align}
\begin{align}
\ket{\psi_{0,0,z}}=&\frac{\alpha\gamma}{\sqrt{2}}\left[\ket{T_0}_c\ket{S}_a\ket{T_0}_t-\ket{T_0}_c\ket{T_0}_a\ket{S}_t\right]+\nonumber\\
&\frac{\alpha\delta}{\sqrt{2}}\left[\ket{T_0}_c\ket{S}_a\ket{S}_t-\ket{T_0}_c\ket{T_0}_a\ket{T_0}_t\right]+\nonumber\\
&\frac{\beta\gamma}{\sqrt{2}}\left[\ket{S}_c\ket{T_0}_a\ket{T_0}_t-\ket{S}_c\ket{S}_a\ket{S}_t\right]+\nonumber\\
&\frac{\beta\delta}{\sqrt{2}}\left[\ket{S}_c\ket{T_0}_a\ket{S}_t-\ket{S}_c\ket{S}_a\ket{T_0}_t\right].
\end{align}

Measuring the ancilla in a singlet or triplet results in the final eight states:
\begin{align}
\ket{\psi_{1,1,1}}=&\alpha\gamma\ket{T_0}_c\ket{S}_t+\alpha\delta\ket{T_0}_c\ket{T_0}_t+\nonumber\\
&\beta\gamma\ket{S}_c\ket{T_0}_t+\beta\delta\ket{S}_c\ket{S}_t,
\end{align}
\begin{align}
\ket{\psi_{1,1,0}}=&\alpha\gamma\ket{T_0}_c\ket{T_0}_t+\alpha\delta\ket{T_0}_c\ket{S}_t+\nonumber\\
&\beta\gamma\ket{S}_c\ket{S}_t+\beta\delta\ket{S}_c\ket{T_0}_t,
\end{align}
\begin{align}
\ket{\psi_{1,0,1}}=&-\alpha\gamma\ket{T_0}_c\ket{S}_t-\alpha\delta\ket{T_0}_c\ket{T_0}_t+\nonumber\\
&\beta\gamma\ket{S}_c\ket{T_0}_t+\beta\delta\ket{S}_c\ket{S}_t,
\end{align}
\begin{align}
\ket{\psi_{1,0,0}}=&\alpha\gamma\ket{T_0}_c\ket{T_0}_t+\alpha\delta\ket{T_0}_c\ket{S}_t-\nonumber\\
&\beta\gamma\ket{S}_c\ket{S}_t-\beta\delta\ket{S}_c\ket{T_0}_t,
\end{align}
\begin{align}
\ket{\psi_{0,1,1}}=&\alpha\gamma\ket{T_0}_c\ket{T_0}_t+\alpha\delta\ket{T_0}_c\ket{S}_t+\nonumber\\
&\beta\gamma\ket{S}_c\ket{S}_t+\beta\delta\ket{S}_c\ket{T_0}_t,
\end{align}
\begin{align}
\ket{\psi_{0,1,0}}=&\alpha\gamma\ket{T_0}_c\ket{S}_t+\alpha\delta\ket{T_0}_c\ket{T_0}_t+\nonumber\\
&\beta\gamma\ket{S}_c\ket{T_0}_t+\beta\delta\ket{S}_c\ket{S}_t,
\end{align}
\begin{align}
\ket{\psi_{0,0,1}}=&\alpha\gamma\ket{T_0}_c\ket{T_0}_t+\alpha\delta\ket{T_0}_c\ket{S}_t-\nonumber\\
&\beta\gamma\ket{S}_c\ket{S}_t-\beta\delta\ket{S}_c\ket{T_0}_t,
\end{align}
\begin{align}
\ket{\psi_{0,0,0}}=&-\alpha\gamma\ket{T_0}_c\ket{S}_t-\alpha\delta\ket{T_0}_c\ket{T_0}_t+\nonumber\\
&\beta\gamma\ket{S}_c\ket{T_0}_t+\beta\delta\ket{S}_c\ket{S}_t.
\end{align}

Applying the gates $\hat{\sigma}_c,\hat{\sigma}_t$ on the control and target qubits in
accordance with the results of the parity and ancilla measurements (as shown in Fig. \ref{gate})
gives the same state in all eight computation branches (up to a global phase):
\begin{eqnarray}
\ket{\psi_f}=&\alpha\gamma\ket{T_0}_c\ket{T_0}_t+\alpha\delta\ket{T_0}_c\ket{S}_t+\nonumber\\
&\beta\gamma\ket{S}_c\ket{S}_t+\beta\delta\ket{S}_c\ket{T_0}_t.
\end{eqnarray}

Under the qubit encoding this state is indeed the result of the CNOT gate:
\begin{eqnarray}
\ket{\psi_f}=&\alpha\gamma\ket{0}_c\ket{0}_t+\alpha\delta\ket{0}_c\ket{1}_t+\nonumber\\
&\beta\gamma\ket{1}_c\ket{1}_t+\beta\delta\ket{1}_c\ket{0}_t.
\end{eqnarray}
This proves the linearity of the gate.

\end{document}